\documentclass[sigconf]{acmart}
\usepackage{booktabs}
\usepackage{xcolor}
\usepackage{multirow}
\usepackage{graphicx}
\usepackage{float}
\usepackage{hyperref}
\usepackage{amsmath,array}
\usepackage{booktabs} % For formal tables

\setcopyright{rightsretained}

\acmConference[SIGKDD'18]{ACM SIGKDD}{August 2018}{London, United Kingdom}
\acmYear{2018}
\copyrightyear{2018}

\begin{document}

\title[Water Main Breaks in Syracuse]{Using Machine Learning to Assess the Risk of and Prevent Water Main Breaks}

\author{Avishek Kumar}
\affiliation{%
  \institution{University of Chicago}
}
\email{avishekkumar@uchicago.edu}

\author{Syed Ali Asad Rizvi}
\affiliation{%
  \institution{University of Oxford}
}
\email{syed.rizvi@eng.ox.ac.uk}

\author{Benjamin Brooks}
\affiliation{% 
  \institution{University of Chicago}
}
\email{benjamin.p.brooks@gmail.com}

\author{R. Ali Vanderveld}
\affiliation{%
  \institution{ShopRunner}
}
\email{ali.vanderveld@gmail.com}

\author{Kevin H. Wilson}
\affiliation{%
  \institution{The Lab @ DC}
}
\email{kevin.wilson@dc.gov}

\author{Chad Kenney}
\affiliation{%
  \institution{City of Denver, CO }
}
\email{ckenney4@gmail.com}

\author{Sam Edelstein}
\affiliation{%
  \institution{City of Syracuse, NY }
}
\email{sedelstein@syrgov.net}

\author{Adria Finch}
\affiliation{%
  \institution{City of Syracuse, NY}
  }
\email{afinch@syrgov.net}

\author{Andrew Maxwell}
\affiliation{%
	\institution{City of Syracuse, NY}
    }
\email{amaxwell@syrgov.net}

\author{Joe Zuckerbraun}
\affiliation{%
	\institution{City of Syracuse, NY}
    }
\email{jzuckerbraun@syrgov.net}

\author{Rayid Ghani}
\affiliation{%
	\institution{University of Chicago}
    }
\email{rayid@uchicago.edu}

\thanks{This work was initiated as part of the 2016 Eric and Wendy Schmidt Data Science for Social Good Fellowship at the University of Chicago.

Any opinions, findings, and conclusions or recommendations expressed in this material are those of the authors and do not necessarily reflect the views of our employers.}

\renewcommand{\shortauthors}{A. Kumar et al.}

\begin{abstract}
Water infrastructure in the United States is beginning to show its age, particularly through water main breaks. Main breaks cause major disruptions in everyday life for residents and businesses. Water main failures in Syracuse, N.Y. (as in most cities) are handled reactively rather than proactively. A barrier to proactive maintenance is the city's inability to predict the risk of failure on parts of its infrastructure. In response, we worked with the city to build a ML system to assess the risk of a water mains breaking. Using historical data on which mains have failed, descriptors of pipes, and other data sources, we evaluated several models' abilities to predict breaks three years into the future. Our results show that our system using gradient boosted decision trees performed the best out of several algorithms and expert heuristics, achieving precision at 1\% (P@1) of 0.62. Our model outperforms a random baseline (P@1 of 0.08) and expert heuristics such as water main age (P@1 of 0.10) and history of past main breaks (P@1 of 0.48). The model is deployed in the City of Syracuse. We are running a pilot by calculating the risk of failure for each city block over the period 2016-2018 using data up to the end of 2015 and, as of the end of 2017, there have been 33 breaks on our riskiest 52 mains. This has been a successful initiative for the city of Syracuse in improving their infrastructure and we believe this approach can be applied to other cities. 
\end{abstract}

\maketitle

\section{Introduction}

\begin{figure*}[t!]
	\includegraphics[width=\linewidth]{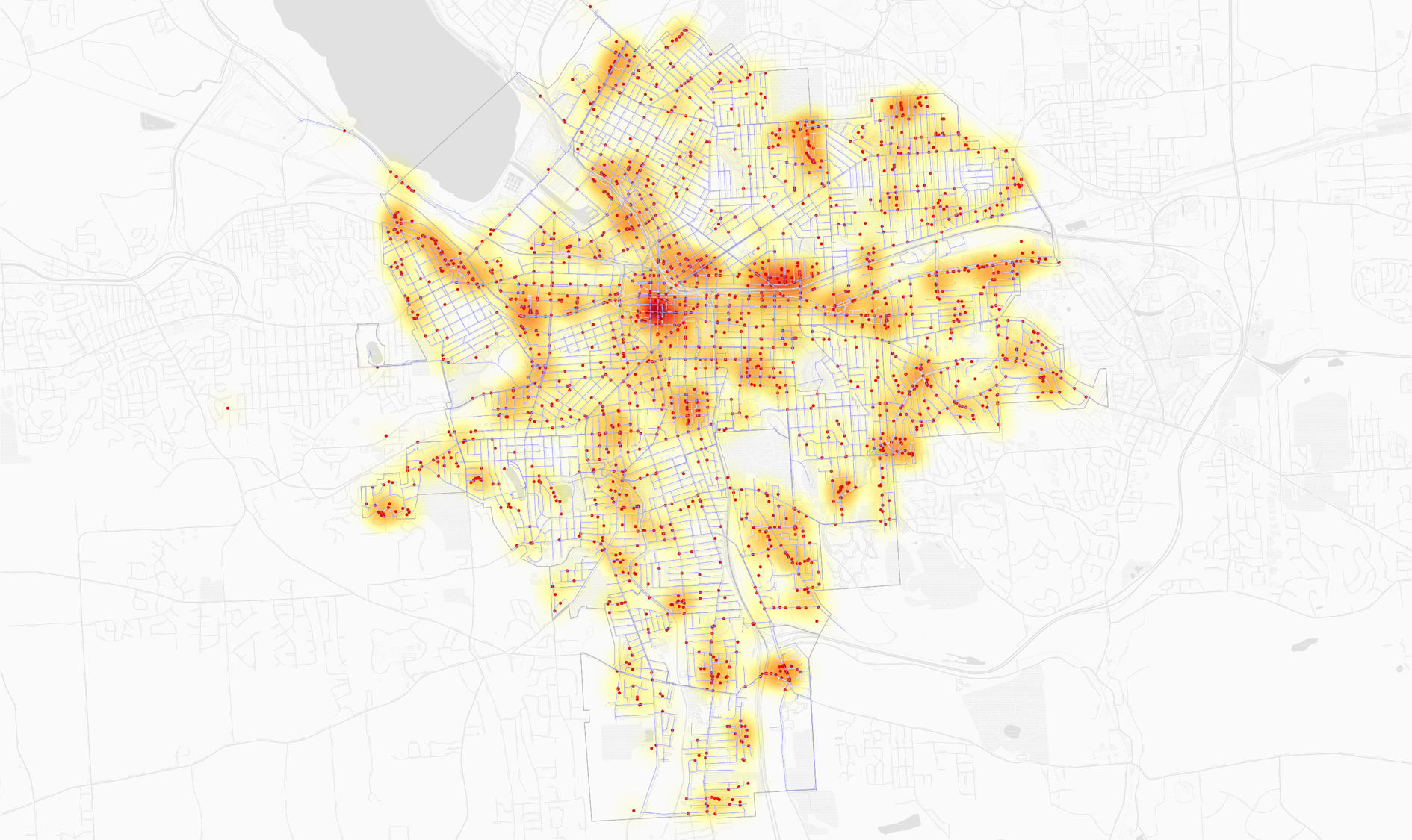}
    \caption{Map of the city of Syracuse. The red dots indicates a main break that occurred between 2004-2016.}
    \label{SyracuseMainBreaks}
\end{figure*}

In 2013, the American Society for Civil Engineers estimated that the United States needed to invest \$3.6 trillion by 2020 to replace its ailing infrastructure \cite{asce_2013_2013}. Neglecting potholes on roads can lead to poor quality roads that become much more costly to fix rather than maintaining roads that are in good condition \cite{anderson_problem_2011}; neglecting bridge maintenance can lead to bridge collapse and death \cite{sander_bridge_2017}; neglecting water treatment can lead to lead from aging service lines to leach into a water supply which subsequently can lead to a city-wide lead poisoning of a generation of children as tragically exemplified in Flint, Michigan \cite{city_of_flint_state_2015}.  Our work concerns the more than one million miles of water mains in the United States that deliver potable water to homes and businesses. The problem of main breaks is particularly prominent in older cities where large portions of the water infrastructure were laid a century or more ago and have passed their operating life. Massive main breaks can lead to property damage by flooding homes and businesses, create massive sinkholes that destroy roads and vehicles on those roads, lead to leaks into gas lines preventing homes from receiving heat, and destroy power lines preventing homes from receiving power \cite{jerome_infrastructure_2017}. In addition to massive main breaks, many cities, like Syracuse, NY, do not suffer from just these massive main breaks. Rather, they are afflicted by a large number of frequent, pestilential, small main breaks. For cities with a growing population it is much easier to launch infrastructure improvement projects or simply cope with expensive reactive maintenance due to a growing tax base. On the antipodal end of the spectrum are cities with a declining population where the costs of maintenance increase while the remaining residents must each bear a larger amount of those costs \cite{semuels_tale_2016}.  
 
\subsection{Syracuse and Water Main Breaks}
The city of Syracuse, NY, has approximately 200 main breaks a year, sometimes suffering multiple breaks in a single day (Figure \ref{SyracuseMainBreaks}). Water main problems are notoriously difficult to monitor given their subterranean location. The city's current strategy (similar to most cities) is reactive - to fix main breaks after they have broken. This purely reactive strategy is untenable due to the increasing cost and number of main breaks and disruption to daily life. The financial resources to handle the water infrastructure has also decreased. Syracuse's population has fallen 35\% from its 1950 peak of 220,000 to 144,000 today forcing current residents to bear the costs of maintaining an aging infrastructure designed for a much larger population. It is therefore critical for cities such as Syracuse to perform preventative maintenance efficiently on their infrastructure as opposed to the reactive process they've been following so far. 

Water main breaks typically occur without any prior warning making it difficult to decide which water mains should be pro-actively replaced before a break.  A water main may break once and not break again for many years while other water mains in different parts of the city break frequently. There are many competing theories about the causes of main breaks from different members of city government. Some believe that the material of a water main is the most important predictive feature of a main break; others believe that the age of a water main is the most predictive feature. Syracuse's industrial heritage had once resulted in the unfortunate designation of Lake Onondaga, located at the city's northwest corner, as ``the Most Polluted Lake in America'' by the New York State Department of Energy and Conservation \cite{MostPollutedLake}. Due to a massive investment and continuous efforts to clean the lake, it has since been cleaned, though the toxic soil remains a concern for water main integrity.  As a result, the soil around and near the lake contains varying levels of toxicity that are thought to contribute to the degradation of water mains.  Additional proposed causes of water main breaks are large amounts of vibrations on roadways above a water main or nearby construction. Competing theories and the lack of an obvious pattern makes the problem of selecting which mains need to be prioritized for replacement nontrivial.
 We focused on the problem of predicting which city blocks in Syracuse are most likely to have water main breaks in the next three years.

\subsection{Existing Approaches} Failure analysis of water infrastructure is a problem that has been studied for the last 40 years using a variety of methods and frameworks often with conflicting results. The first known paper is \cite{shamir_analytic_1979}. In that work the authors predicted the number of failures per unit length using a Poisson model based on the age of the pipes. There has also been a long history of modeling water main breaks using survival-based methods such as homogeneous Poisson process, Poisson regression \cite{asnaashari_prediction_2009,boxall_estimation_2007}, exponential Weibull model and Weibull hazard model \cite{vanrenterghem-raven_statistical_2003,le_gat_using_2000} . Oftentimes these models separate water mains into classes based on material and diameter and calculate the failure rate as a function of time. There are also conflicting results where various predictive variables are thought to play a factor. For instance, pipe age is the main issue in \cite{shamir_analytic_1979} and soil resistivity is the main risk factor in another model \cite{kabir_predicting_2015}, whereas number of breaks can be the main predictive factor in other models \cite{pelletier_modeling_2003,le_gat_using_2000,mailhot_modeling_2000}. Review articles on failure analysis can be found in \cite{scheidegger_extension_2013} and \cite{kleiner_comprehensive_2001}. 

\subsection{Our Contribution}
We framed this problem as a binary classification problem of whether a water main break will occur on a given city block within the next three years. This definition allows the city of Syracuse to easily operationalize this model and plan the infrastructure development for the next three years. We compare our results to a variety of baseline approaches including heuristics that are currently used by experts in this area. Our publicly available code at \url{https://github.com/dssg/syracuse_public} allows other cities to replicate this analysis and compare it with existing approaches they are using. Our contribution to the problem is a model that uses out-of-sample prediction, temporal cross-validation, and tunes the model to the capacity of the city to intervene. Our results on historical data showed that we can accurately estimate the probability of a water main break and are 62\% accurate in the top 1\% of our predictions. Based on those results, the city of Syracuse has deployed this system and is currently conducting a field trial to validate these predictions over the next year. At the time of writing there have been 33 breaks among our top 52 predicted city blocks. 

The city of Syracuse currently plans to use this system in two ways: 1) for preventative maintenance on the top 1\% of the riskiest water mains 2) to use the risk scores to coordinate with the Department of Public Works (DPW) as they do road construction and maintenance. This has not only been a successful initiative for the city of Syracuse in improving their infrastructure but we also  believe this approach can be successfully applied to other cities around the world. Our code has been released as open source and is available on github at \url{https://github.com/dssg/syracuse_public} for other cities to reuse and extend. The rest of the paper describes the problem in more detail, our solution, and validation results.

\section{Water Main Problem in Syracuse, NY}
\begin{table}
\centering
\begin{tabular}{|p{2.5cm}|p{1cm}|p{1cm}|p{1cm}|}
\hline
Street & Block & Road Rating & Risk Score \\
\hline
E Hampton Ave & 900-934 & 8 & 81 \\
\hline
Smith St & 400-438 & 0 & 78 \\
\hline
S Hardy Dr & 200-299 & 4 & 74 \\
\hline
Durham Ave & 400-499 & 7 & 73 \\
\hline
Roanoke St & 800-899 & 6 & 70 \\
\hline
\end{tabular}
\caption{An example of the output of the model. The city can use the combination of
risk score, a measure of how likely there is to be a water main break on a block, and the 
road rating, a rating from 1-10 on the quality of the road, to plan infrastructure upgrades.}
\label{table: output}
\end{table}

\begin{figure}[h]
	\includegraphics[width=\linewidth]{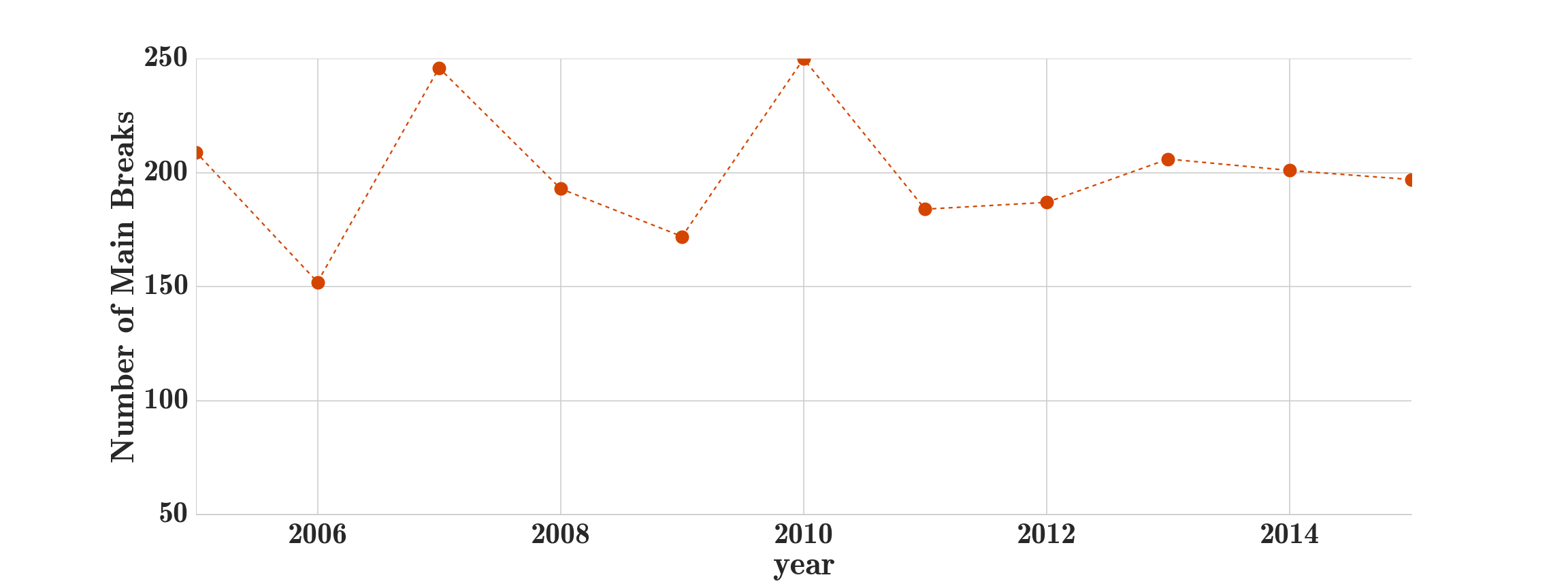}
    \caption{There are approximately 200 water main breaks every year in the city of Syracuse.}
    \label{FIG:breaks_per_year}
\end{figure}

\begin{figure}[h]
	\includegraphics[width=\linewidth]{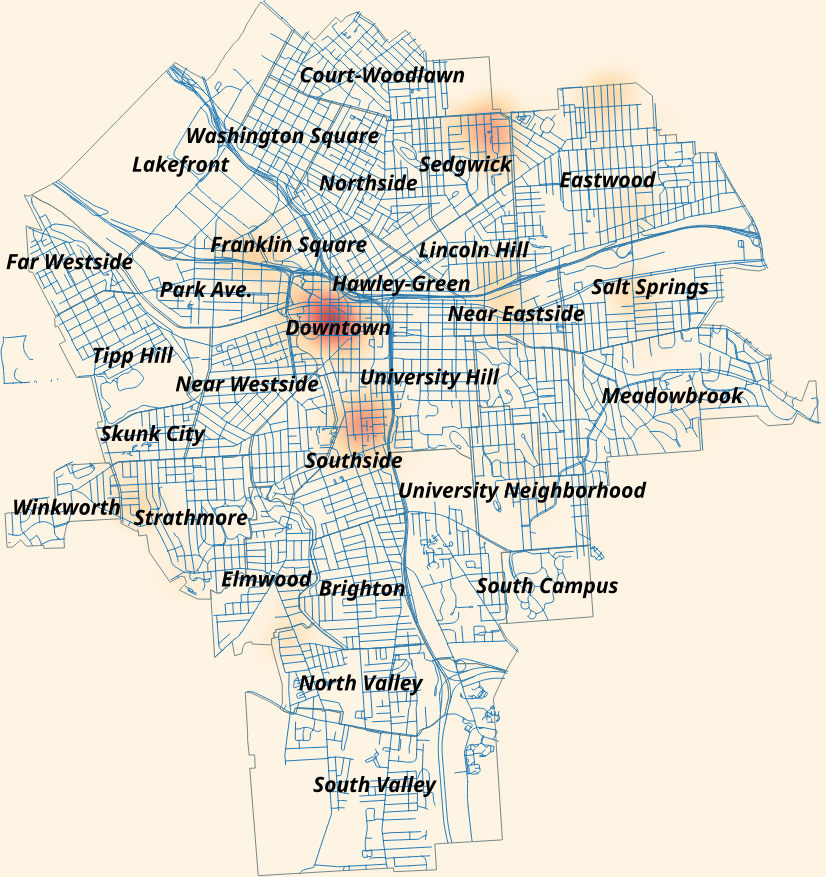}
    \caption{Spatial clustering indicates that clusters are localized in regions where there are multiple main breaks typically on the same water main; the Downtown, Southside and Sedgwick areas in particular contain "hotspots" where water mains have frequent breaks.}
    \label{FIG:syracuse_clustering}
\end{figure}

The city of Syracuse experiences approximately 200 water main breaks a year with the most frequent water main breaks occurring in the winter months( Figure \ref{FIG:breaks_per_year} ). Exploratory spatial clustering \cite{GEAN:GEAN912,GEAN:GEAN261} (Local Getis-Ord Clustering) of main breaks reveals that clusters are made up of small localized clusters with frequent water main breaks typically on the same main particularly in the Downtown, Southside and Sedgwick neighborhoods of Syracuse (Figure \ref{FIG:syracuse_clustering}). Issues with the water main system in the Downtown area are largely attributed to the age and material of the pipes; the Southside and Sedgwick areas are known to have issues with water pressure. 

This dispels the notion that there are large regions of the city where there are predominantly more main breaks than other regions. For instance it was believed that due to Syracuse's industrial heritage the area near Lake Onondaga has a greater frequency of main breaks than other other parts of the city due to soil toxicity.  

Each main break in the city of Syracuse requires dispatching a repair crew to dig up the main and weld a piece of iron over the break, disrupting water service to residents and businesses. Moreover, the quality and lifespan of the road above the broken main is significantly degraded. While the city's Water Department is responsible for maintenance of the system, repairs and maintenance of the water mains necessitates coordination with the Department of Public Works (DPW), which maintains the roadways above the water mains. DPW currently conducts an annual survey of its roads to identify those most in need of repair or replacement, the Water Department had little means of identifying mains at the highest risk of replacement besides relying on the expertise of field crews who deal with main failures on a daily basis.

Our analysis is facilitating planning and coordination -- by providing the Water Department with an accurate risk assessment about their system such that they are able to conduct targeted proactive main replacement in areas that align with DPW priorities. For instance, if a city block is likely to have a water main break in the near future that needs a road replacement, it would be best for the city to replace the water main on the block before replacing the road. This prevents a water main break from destroying a newly replaced road stretching Syracuse's limited resources further, table \ref{table: output} is an example of the output provided to the city. 

Our model is deployed and we are currently conducting a field trial. The risk scores for each city block are used in an asset management system by the water department.  The risk scores and road ratings have also been used to inform the city's five-year plan for upgrading the city's road and water main infrastructure.   

\section{Data Sources and ETL}
We combined and digitized several disparate data sources including 100-year-old field notebooks on the layout of pipes, excel spreadsheets of past work orders, historical property tax records, geological data, and shapefiles of the water system and roads layout of the city to create our dataset.

\begin{figure}[h]
	\includegraphics[width=\linewidth]{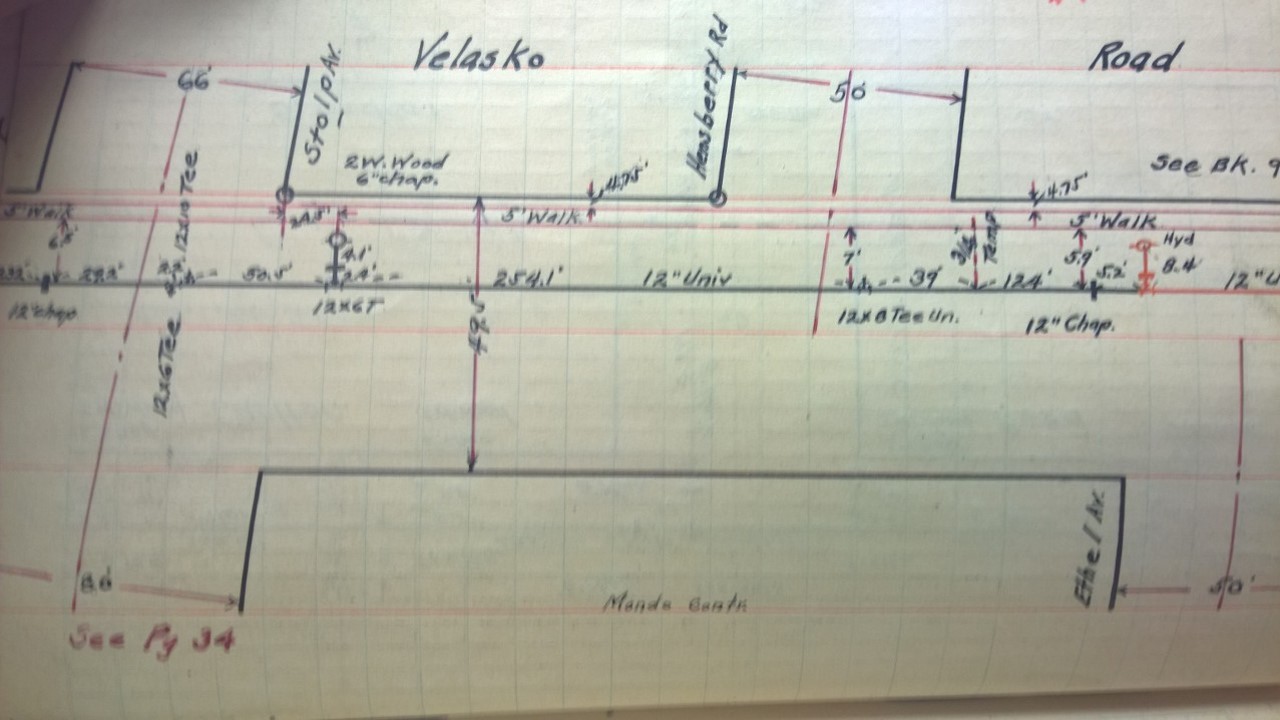}
    \caption{A hand drawn map of the water main layout on Velasko Road in the city of Syracuse from an original field notebook.}
    \label{FieldNotebook}
\end{figure}

\begin{table}
\centering

\label{table:1}

\begin{tabular}{|p{4cm}|p{4cm}|}
\hline
Database & Type \\
\hline
Water Department Work Orders & Records of work done by the water department \\
\hline
Water Main System & GIS Shapefile of the water infrastructure \\
\hline
Tax Parcel Data & Historical property tax records \\
\hline
Field Notebooks & Hand drawn field notebooks of water main layout \\
\hline
Road Ratings & Ratings of the quality of roads in Syracuse \\
\hline
Soil Data & GIS Shapefile on soil type and rocktype provided by the city of Syracuse \\
\hline
County Shape File & GIS Shapefile of all roads and city blocks in Syracuse \\
\hline
\end{tabular}
\caption{Types of data used to create the model}
\end{table}

\subsubsection{Water Department Work Orders}
This data contains a list of services provided by the Department of Water on the water system from 2004-2015 (approximately 10,000 rows of data). This includes a job description, the date of work, and address.  We used each address to map a reported break to the specific water main where the break occurred.

\subsubsection{Water Main System}
The city has the location of all water mains, hydrants, valves, reservoirs and pressure zones as shapefiles. These shapefiles were created by Syracuse's Water Department by digitizing old field notebooks and recording new construction. The shapefiles also contain some information on pipe diameter, pipe material, year of installation, soil data and pressure data. There are approximately 500 miles of water mains in the city, resulting in about 36,000 rows of data.  

\subsubsection{Tax Parcel Data}
This is a shapefile containing historical property tax records for every land parcel in the city. 

\subsubsection{Field Notebooks}
Field notebooks have data on pipe age, material, and diameter that were recorded by the original field engineer at the time of installation. This was used to find a large number of pipe ages and materials as explained in the methodology section.  

\subsubsection{Road Ratings}
We received road rating data from 2000-2015. All roads are rated from 1 to 10, with 10 being a perfect newly paved road and 1 being a dirt road. It is possible there could be a correlation between poor roads and water main breaks. We did not find one, nor did the road ratings help our predictions. Given street addresses we were able to geocode the road ratings. 

\subsubsection{Soil Data}
Soil data was obtained from the USGS database in the form of a shapefile. Syracuse contains 55 distinct types of soils which may play a role in whether water main breaks occur or not.  

\subsubsection{Syracuse Streetline Shape File}
The county shape file in Syracuse maps all the roads and individual city blocks.   

\subsubsection{Limitations of the Data}
The data sources described above have several limitations.  For instance, there are several ways that water mains break. There are ``round cracks'' associated with the depth of the frost line in winter or drop in water temperature, ``spilt cracks'' associated with the expansion of metal in the summer, and wholesale breaks where entire streets collapse. Despite wide ranging severity, these are all labeled ``Main Break/Leak'' in our data. Moreover, there is a great deal of lore around why pipes break. Soil toxicity, soil composition, nearby jackhammering, road construction and the (de)activation of reservoirs feeding the water system were all proposed as stressors on mains. Unfortunately, data on these events and descriptors were either unavailable or not sufficiently robust to be used in our model. 

\section{Methodology}

All of the code used in this paper may be found at \url{https://github.com/dssg/syracuse_public}. This section describes our overall methodology, from extracting features to building machine learning models to model selection and evaluation.

\subsection{Feature Extraction}
Our unit of prediction (row of data in training/testing set) is a city block in Syracuse. The block level was chosen after discussions with different departments within the city of Syracuse: when replacing a water main, the water main along an entire city block would be replaced. The features of our model were diameter, pipe age, installation year, soil type, rock type, pressure zone, pipe material, road rating, number of previous breaks on that city block, number of previous breaks on city blocks nearby. Generally, the diameter, pipe age, installation year, soil type, rock type, pressure zone and road rating were the same across a city block. The features created in our system were based on expert knowledge and a thorough examination of the data. Historical studies pointed toward pipe material, pipe age, soil type, toxicity level of the soil, geological composition, and topography as potential factors that affect the lifespan of water mains \cite{morris_principal_1967,shamir_analytic_1979,kabir_predicting_2015,pelletier_modeling_2003,le_gat_using_2000,mailhot_modeling_2000,le_gat_using_2000}. The Syracuse water main system is largely gravity driven. There are several parts of the city for which varying topography leads to different pressure zones. City officials believe that a pressure zone can play a role in main breaks. We had to impute many of these features due to a lack of coverage in our data.

The water main shape file maintained by the city only contained 2\% coverage of pipe age, diameter and material. To impute pipe age, material and diameter we had to use the original handwritten field notebooks. An intern with the city of Syracuse digitized, by hand, the relevant notebooks that contained hand-sketched drawings (Figure \ref{FieldNotebook}) of the water main system using a sample strategy developed through conversations with the city: First, we used historical property tax data to find the year a home on a city block was assessed property tax; this year was used as a substitution for the year a pipe was installed on that block. Once we had large coverage on the installation year of the pipes, we used a set of rules to determine the pipe material. If a pipe was installed before 1920 we assumed Cast iron; if a pipe was installed after 1960 we assumed the pipe is ductile iron; for all pipes installed between 1920-1960, we looked up the pipe material and diameter in the original field notebook. In the cases where the shape files contained pipe age and material we also looked up the pipe diameter. We were able to make the simplifying assumption that we could find the pipe age, material and diameter for a single block on a street and then apply these values to all blocks on that street due to the homogeneity of water mains along a street.    

\subsection{Aggregation of water mains to city block level}
The location of each water main was mapped to a city block using PostGIS. Due to distortions in the projections in the street line shapefile and water main shapefile we could not simply map a water main to street based on whether a water main line was aligned to a street line. Instead, we created a buffer around the street lines using the PostGIS buffer function and then calculated the maximum overlap a water main line had within a buffered street. A mapping was made between a water main and street block when the water main line fell in the area inside the buffered area of that street block. By examining the number of breaks on a given city block it became apparent that the more water main breaks on a city block the greater likelihood for a water main break to occur on that block again. We, therefore, incorporated the number of past main breaks on a city block as a feature. All categorical features were converted to dummy variables. 

In our data, the water mains were recorded as GIS line segments that began or ended between fire hydrants, valves, or, sometimes, arbitrary points. However, the natural unit for proactive repair is not a GIS line segment, but a city block due to the block level being the level at which there would be an intervention, which, in this case, is the replacement of a water main on a city block. Thus, we aggregated our data to this resolution to enable preventative maintenance by the city. In the rare instances when pipe age, material and diameter were not constant along a city block, we took the earliest install year, majority material by length, and smallest diameter. 

\begin{table}
\centering
\begin{tabular}{|p{3.5cm}|p{4.5cm}|}
\hline
Feature & Description  \\
\hline
diameter & Diameter of water main  \\
\hline
pipe age & Age of the water main at the time of prediction  \\
\hline
installation year & year water main installed \\
\hline
soil type & Soil type from USGIS   \\
\hline
rocktype & Rock type from USGIS  \\
\hline
pressure zone & The city of Syracuse's water system is largely gravity driven. In parts
                of the city there are different pressure zones that are used to maintain the
                water system.\\
\hline
material &  Material the the water main is made of \\
\hline
road rating of block & A rating from 1-10 on the quality of the road above the main. \\
\hline
number of previous breaks & The number of previous water main breaks on that city block. \\
\hline
number of breaks nearby & The number of previous water main breaks in a 100ft radius. \\
\hline
\end{tabular}
\caption{List of features used by the model}
\label{table: features}
\end{table}

\section{Modeling Approach and Results}
The problem was cast as a binary classification problem to predict which mains are likely to break in the next three years so they can be proactively replaced before the break. The three-year prediction window and precision at 1\% were based on the resources the city has to replace water mains. There are 5263 city blocks in Syracuse; the city can reasonably replace water mains of 52 blocks every three years, making precision and recall at 1\% (52 blocks) the metric to optimize.  Models were evaluated using temporal validation\cite{hyndman_forecasting:_2014} where a model is trained on data from a particular time period and evaluated on the next time period. For example, if we are assessing the risk of a main break between the years 2013 and 2015 (break within 3 years), we train our models using features from before 2013 and test them on the three years starting form 2013. With the data ranging from 2005-2015, multiple training and test sets were created and the results aggregated for model selection. For each test set, we calculated  precision at the top 1\% for predicting breaks three years out in the future. 

Different classification methods (Decision Trees, Logistic Regression, AdaBoost, Random Forests, Gradient Boosted Decision Trees), training data history, and feature sets where compared to each other using precision at 1\%. In addition, machine learning models where compared to several baselines: a random baseline as well as a few based on heuristics that domain experts use, pipe age and number of past breaks. We wanted to compare the machine learning models we were building to each other, to a random baseline, as well as to the set of heuristics that experts within the water infrastructure community would use today to predict breaks to make sure the system is in fact an improvement and worth deploying.  To assess how well the models estimate the probability of a main break, we used decile analysis to compare the predicted probability to the empirical probability.

\subsection{Predictive Performance}
The model that performed the best at precision at 1\% used the Gradient Boosting Decision Trees (GBDT) algorithm looking back six years into the past. GBDT constructs an additive regression model which uses a linear combination of decision trees \cite{friedman_stochastic_2002}. GBDT can perform feature selection inherently and the number of boosting iterations is a natural regularization parameter \cite{friedman_stochastic_2002}. In the GBDT classifier, the model follows an additive expansion of the following form:

\begin{equation}
f(X,\beta,\alpha) = \sum_{j=1}^{n}\beta_{j}h_{j}(X,\alpha_{j})
\end{equation}

where $h$ are the base learners, $\beta _j$ is the weight of tree $h_j$, $\alpha _j$ denotes the parameters of the $j$th decision tree $h(X,\alpha _j)$, splitting variables, split locations, and the terminal node means of the individual trees, X is the feature vector. The mean loss function is defined as 

\begin{equation}
Obj^{(j)} = \sum_{i=1}^{n}(y_{i} - (\widehat{y_{i}}^{j-1} + h_{j}(X,\alpha_{j}))^{2})
\end{equation}

Where $y_{i}$ is the ground truth, $\widehat{y_{i}}^{j-1}$ is the score from the j-1 iteration and $h_{j}(X,\alpha_{j})$ is the j-th iteration base learner. 

The boosting algorithm in GBDT is the following:

\begin{itemize}
\item Initialize the list of base learners with a simple uniform weight assigned to each base learner.
\item In the first iteration check which training examples $(X,y)$ were poorly predicted by the forest.
\item Increase the weight of the examples that the existing forest predicted poorly.
\item Estimate a new weak classifier $h_i$ based on the weighted examples.
\item Compute the weight $\beta_i$ of the new week classifier.
\item Add the new classifier $(h_i, \beta_i)$ to the forest.
\item Repeat iterations by re-weighting the training examples and adding further weak classifiers.
\end{itemize}

\begin{figure}[h]
	\includegraphics[width=\linewidth]{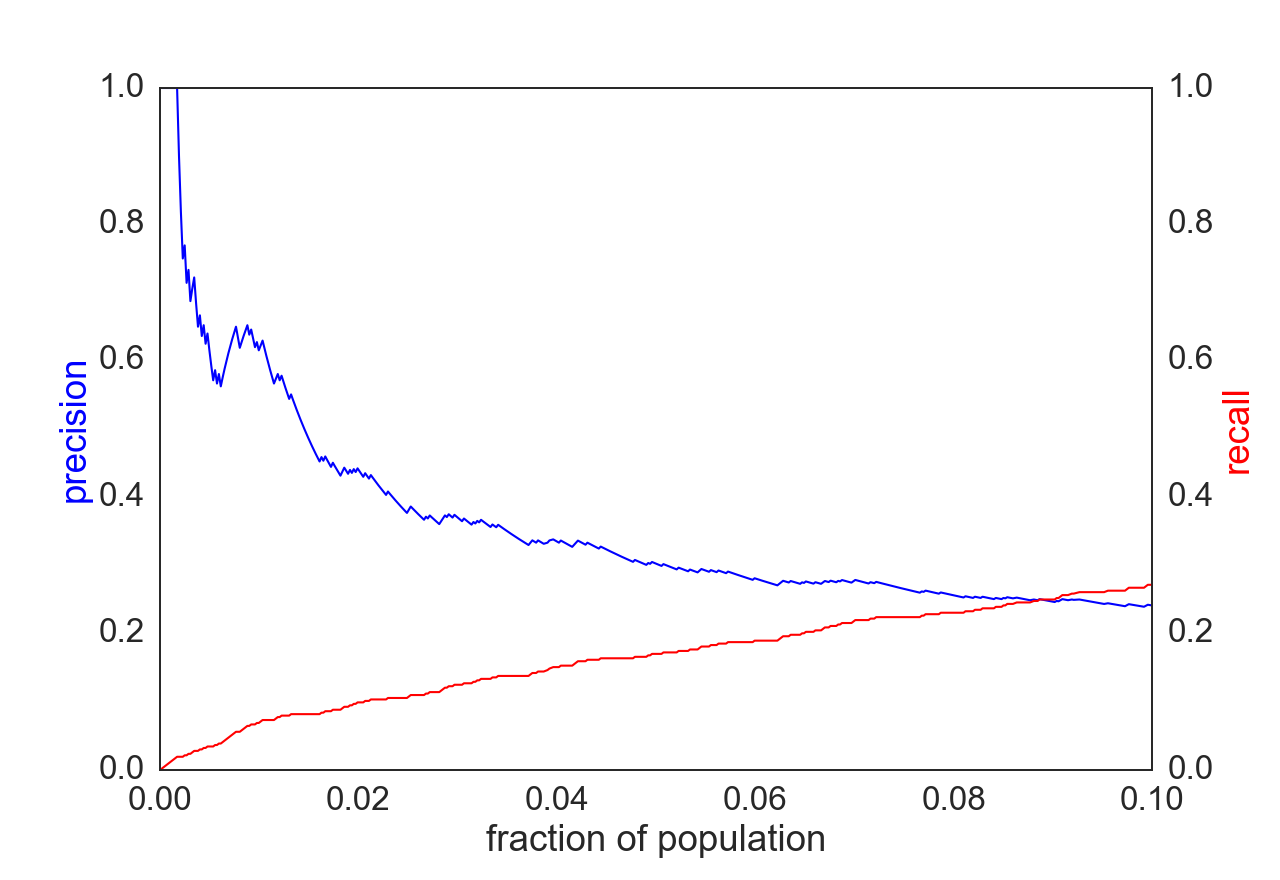}
    \caption{ Precision and recall curve using a Gradient Boosting Decision Tree Algorithm. The precision at 1\% (52 city blocks)  is 0.62 (32/52 blocks) and recall at 1\% is 0.07.  
}
    \label{p_n_r}
\end{figure}

Using GBDT we obtained the best performance when the number of boosting iterations was set to 100, and maximum depth of each tree was set to 3 (this in turn imposes a limit on the terminal nodes). We used a subsampling rate of 50\%, meaning that each base learner could use 50\% of the training samples. This subsampling effectively results in Stochastic Gradient Boosting\cite{friedman_stochastic_2002}. The precision at 1\% (52 city blocks) using the best performing GBDT is 0.62 and recall at 1\% is 0.07 (Figure \ref{p_n_r}). This translates into 32/52 city blocks that we correctly predicted would break within three years of the end of the training data.

\subsubsection{Comparison to Baselines}
We compared our machine learning models to  a random baseline and the expert heuristics of pipe age and number of past breaks. A random baseline where 52 city blocks are chosen at random results in a precision at 1\% of 0.08. When ranking all city blocks by pipe age, the precision at 1\% is 0.10, which is close to random. This dispels the common notion that the oldest pipes are likely to fail first and indicates there are a number of other factors involved such as material, soil, pressure zone.  Creating a ranked list of city blocks based on the number of past breaks leads to a much better precision at 1\%, 0.48, compared to pipe age and a random baseline. This shows that the number of past breaks is a strong but not perfect indicator of water main failure. All our models with the exception of a basic decision tree model, outperform the random baseline and expert heuristics.  The best model outperforms the random baseline and expert heuristics with a precision at 1\% of 0.62, a 14 percentage point improvement. 

\begin{table}
\begin{tabular}{|p{4cm}|p{2cm}|}
\hline
Heuristic/Model & Precision at 1\% \\
\hline
Random (baseline) & 0.08 \\
\hline
Rank by Pipe Age (baseline) & 0.10 \\
\hline
Decision Trees & 0.41 \\
\hline
Rank by Past Breaks (baseline) & 0.48 \\
\hline
AdaBoost & 0.53 \\
\hline
Logistic Regression & 0.58 \\
\hline
Random Forest & 0.60 \\
\hline
Syracuse Model (GBDT) & 0.62 \\
\hline
\end{tabular}
\caption{ Comparison of models with one another, a random baseline and expert heuristics.}
\end{table}

\subsection{Feature Importances}
\begin{figure}
	\includegraphics[width=\linewidth]{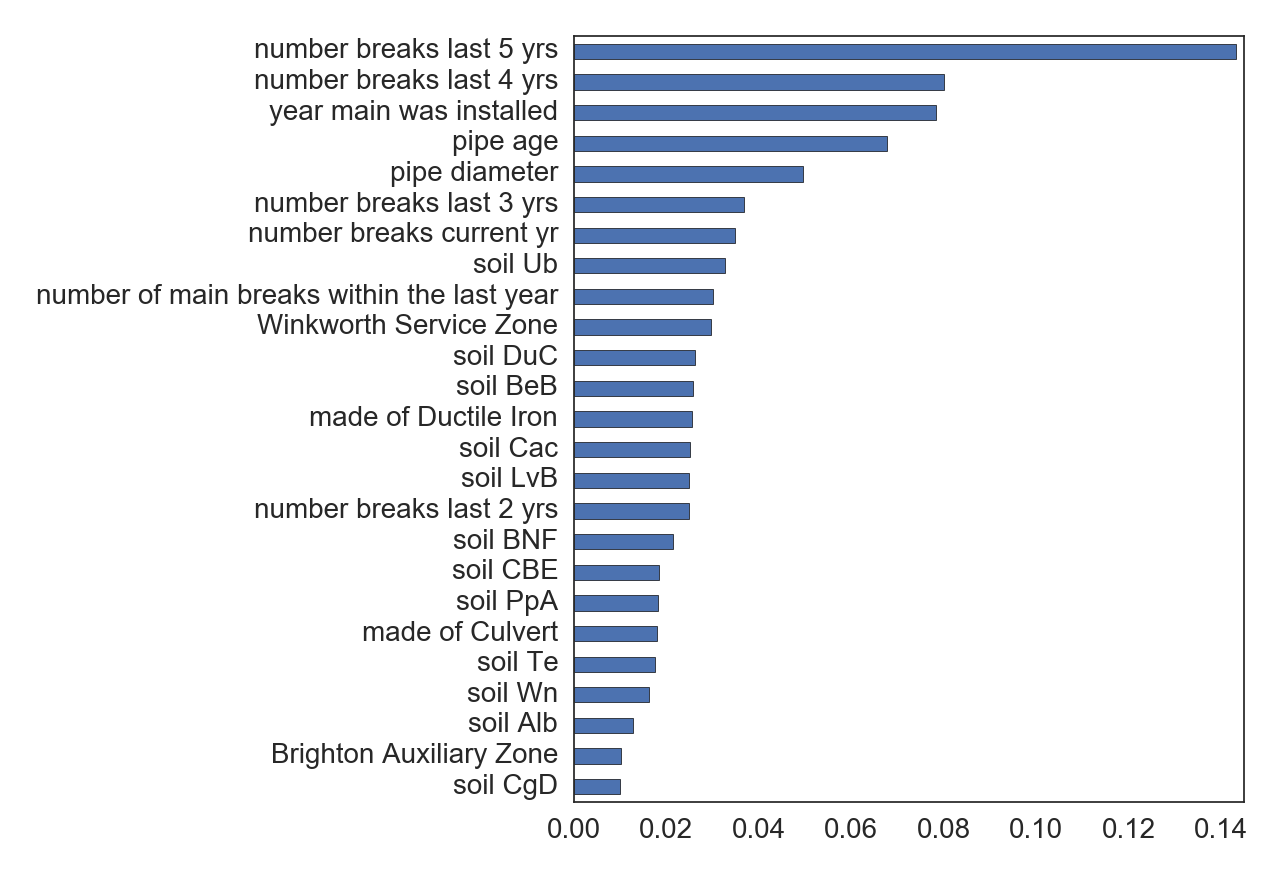}
    \caption{Feature Importances of the best performing model calculated using Gini importance.}
    \label{feat_important}
\end{figure}

Gini importance \cite{Louppe:2013:UVI:2999611.2999660,louppe2014understanding} was used to calculate the feature importances (Figure \ref{feat_important}). The metric is calculated by summing the gain in purity over the number of splits in a tree that use the feature.  As expected, the most important feature of our model is the past history of breaks on a city block with the number of breaks within the last 5 years being the most predictive feature. If a water main continuously breaks every year it is likely that water main is already highly defective, corroded and in need of a replacement. The typical water main repair is to weld a piece of iron over the cracked portion of the water main rather than replacing the water main due to costs and construction time; the repair does not fix underlying structural issues. Pipe diameter is also an important feature. As our experience suggests, the larger the diameter a pipe has the more resilient it will be to temperature changes and a shifting of the ground due to traffic or the bending and shifting of soil. 

Despite pipe age not being a sole useful heuristic in predicting main breaks, it is quite important in our model. As pipes age they are likely to fail due to corrosion and normal wear. The interaction of pipe age with the other features in our feature set appears to be predictive water main breaks. The quality of the roads above a water main as captured by the road ratings was not predictive of a water main break. Based on expert knowledge, pipe material was largely thought to play an important role. Throughout the history of the city, installed water mains have been made of several materials. Cast Iron and Universal pipes have long been considered by the water department to be inferior to Ductile Iron and prone to water main breaks.  Surprisingly, pipe material is not an important feature of our model. We also trained models with and without the installation year and interestingly found that using the installation year improved the model. Installation year distinguishes pipes further than simply using age, material and diameter. Pipes made of the same material can be of different quality due to being manufactured by different vendors that were purchased and installed by the city at different times. The model can distinguish types of pipes that are of the same material by using a combination of pipe material and the year the pipe was installed. 

\subsection{Probability Calibration}

\begin{figure}
	\includegraphics[width=\linewidth]{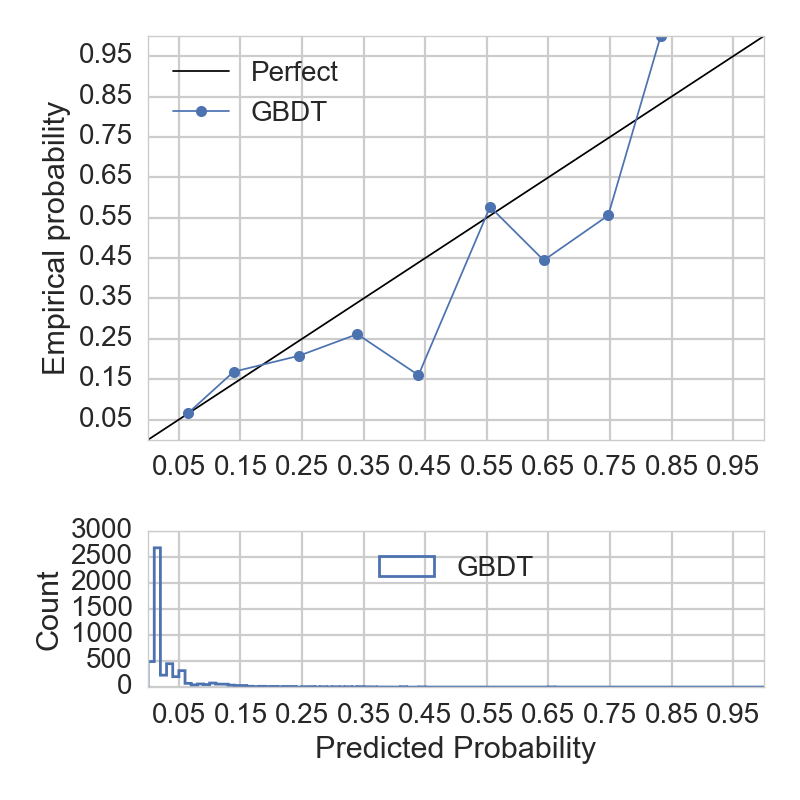}
    \caption{(Top) The reliability curve shows that the probability of a main break is well calibrated. (Bottom) The majority of probabilities are zero as would be expected due to a main break being a rare event, only 8\% of city blocks have experienced a main break in the last 10 years. }
    \label{reliability_curve}
\end{figure}

In order to evaluate if the risk scores can be reasonably used as probabilities to make maintenance decisions, we have to assess how accurate the model's risk scores are to the overall probabilities of a main break. We compared the predicted probability of the model to empirical probabilities using decile analysis\cite{niculescu2005predicting}. A reliability curve was calculated by partitioning the predicted probabilities into ten bins. The mean empirical probability and mean predicted probability for each bin was computed. For perfectly calibrated predictions, the mean empirical and predicted probabilities would be equal. We have also plotted the reliability curve for each decile (Figure \ref{reliability_curve} Top). From the curve we can see the probabilities are mostly close to zero as would be expected due to a main break being a rare event (Figure \ref{reliability_curve} Bottom); typically, only 9\% of city blocks have a water main break over a three-year period. The probabilities are well calibrated in the lower deciles as can be seen from the reliability curves. The middle deciles have predicted probabilities that are slightly greater than the empirical probabilities. The top decile has a predicted probability less than the empirical probability. This is likely due to a lack of samples since the top decile only contains six samples. 

\section{Implementation and Next Steps}
As part of the deployment process, we took our best performing model and retrained it on all the data up to 2016. This retrained model was then used to make predictions on which water main breaks will fail from 2016-2018 - the next three years. 

The city is currently monitoring these predictions as part of a field trial by monitoring the number and location of breaks that occur within this period in order to validate the model. At the time of this writing there have been 33 main breaks on our top 52 predicted city blocks. These risk scores are being used as input in a 5-year plan to upgrade the entire infrastructure of the city\cite{syracuse_i-team_city_2016} and perform proactive maintenance. 

The model has been used in several different ways. First, as the City has tested water sensors that would detect leaks in real time, we used the model to dictate where sensors should be placed. Given that we suspected the mains from the model to be the most likely to break, getting regular updates on the status of those mains was important. The result was that we did identify leaks and other issues with the mains based on those leak sensors. More significantly, the model was used to help give context to blocks of road where all of the infrastructure was in need of replacement. Looking at road and sewer quality, in combination with the riskiest mains from the model, the City was able to justify several Dig Once projects where all infrastructure was replaced at one time; a dig once project being a coordinated effort between various agencies in the city to replace water mains, gas lines, telecommunication lines (full underground reconstruction) and roads such that the city only needs to "dig once" for maintenance of the infrastructure. Additionally, when there has been development in the city and buildings were being restored or built, if the City knew a water main was at risk of breaking, we would encourage the developer to replace the water main at the same time as they were doing other work. These types of projects have saved the city more than \$1.2 million dollars.

Through this project we have found data quality and communication is critical to the success of data science projects tackling urban infrastructure problems. Other cities that are interested in deploying a similar model may not have the same types of data that were used in the project, but it is likely they still have data that can be operationalized. It is also important to properly communicate the output and benefits of the model to stakeholders around the city. Simply asking front-line staff to trust a data model will not convince a department to change its ways, but showing how the model can benefit the department, help make jobs easier, and ultimately save money will lead to success. Our approach enables Syracuse and similar cities to identify city blocks that are most likely to have main breaks and replace water mains proactively, decreasing the number of main breaks in the long-term. If this approach were implemented over an extended period of time, a city can proactively use its limited resources to gradually modernize a city's infrastructure rather than reactively repairing water main breaks that have added costs to residents and businesses and are, in general, disruptive to everyday life. We hope this model will serve as a proof-of-concept for evaluating risk in all infrastructure systems in the city.  

\section{Conclusions}
The present work uses a machine learning approach to develop a risk model for predicting which city blocks are most likely to have a water main break within the next three years. Our model significantly outperforms a random baseline and expert heuristics. The model outputs risk scores, allowing the city to proactively maintain the city's water infrastructure rather than resorting to purely reactive maintenance. The risk scores of the model enables the city to better allocate resources, facilitate long-term planning, and increase cooperation between different departments such as the water department and department of public works.  

The model also provides insight into which factors are important in predicting whether a city block will have a water main failure. We find that largely intuitive features such as the number of past main breaks, pipe age, and pipe diameter, pipe material are important in predicting a main break. However, other common scapegoats such as road quality and road paving seem not to be predictive. It also worth noting that the number of main breaks previously on a block in not perfectly predictive of future main breaks, nor is the age of a water main.   

At a higher level we would like to deploy this model to other cities. The model has been released as an open-source package for other cities to use and extend (\url{https://github.com/dssg/syracuse_public}). Our model is deployed. We are currently monitoring the number and location of main breaks to validate our three-year predictions in our field trial. The risk scores for each city block are used in an asset management system by the water department.  The risk scores and road ratings have also been used to inform the city's five-year plan for upgrading the city's road and water main infrastructure. Ultimately, we hope our work at the intersection of data science and city planning will lead to more proactive infrastructure maintenance. 

\section{Acknowledgements}
We acknowledge funding from the Schmidt Family Foundation through the Data Science for Social Good Fellowship at the University of Chicago. We also thank David Robusto for digitizing the original field notebooks, as well as the Syracuse Water Department and Department of Public Works for sharing data, expertise, time, and feedback. 

\bibliographystyle{ACM-Reference-Format}
\bibliography{sigproc}

\end{document}